# The SETI Paradox

**Alexander Zaitsev**

Institute of Radio Engineering and Electronics,

Vedensky Square 1, Fryazino, 141120 Russia

**Abstract:** Two opposing tendencies paradoxically coexist in terrestrial consciousness – the insistent quest for intelligent signals from other civilizations and the persistent aversion to any attempts to transmit such signals from Earth toward probable fellow intelligent beings. If typical for our entire Universe, such manifestations of intelligence would make the search for other civilizations totally meaningless.

### 1. Introduction

Searching the Internet for the word combination "SETI Paradox" yields two separate and interrelated groups of results – SETI and Fermi Paradox. Here, we focus on the "SETI Paradox" – on this incomprehensible hope of finding extraterrestrial intelligence while keeping almost absolutely silent. However, nothing but natural objects can be found in a Universe where there are only "searchers" and no "senders".

Of the three components of the classical triad "Universe, Life, Mind" that Shklovskii (1962) introduced into scientific and public use, we can now say nothing definite about mind and its possible variety or, on the contrary, sameness. We can only formulate various hypotheses, like, for example, Arthur C. Clarke, who said: "…it is almost evident that biological intelligence is a low form of intelligence. We are at the early stage of the evolution of intelligence, but at the late stage of the evolution of life. True intelligence is unlikely to be living."

The planetary consciousness of the Earth may well be unique and so may be the planetary consciousness of each extraterrestrial civilization. And all planetary consciousnesses in their global, mature manifestations – both internal and external – may well be dismally monotonous, and this very fact may explain the Great Silence – because a passive/receive-only attitude toward the Cosmos is perhaps everybody's, and not just our, feature – every-

body tries to receive and nobody is willing to give…

We suggest introducing – in addition to such common terms as ETI = Extraterrestrial Intelligence and SETI = Search for ETI – a new term, METI = Messaging to ETI, which we use to designate the fundamentally new type of human activities – transmission of messages to hypothetical fellow intelligent beings. Some may argue that SETI is also a new type of activity. Of course, it is a new one, but not fundamentally new – mankind has always been looking into the sky in the hope of finding something there. And as for transmitting to probable ETI and doing this purposefully – this type of activity is now only at its first stages (Zaitsev, Chafer, Braastad, 2005) and it is by no means clear whether it has any future at all…

Shvartsman writes in his already classic paper, "Search for Extraterrestrial Civilization – A Problem of Astrophysics or of the Entire Culture?" (1986):

*"…we do not know for the sake of what transmissions are to be made…"*

and

*"…science is an activity aimed at acquiring new knowledge about the world. However, the interstellar messages are by no means meant to obtain new knowledge by those who transmit them (message and reply are typically several thousand years apart)."*

Indeed, why should we transmit a message to Others? It is more or less clear why we should search for the messages of Others. But why transmit? What for? Indeed, Shvartsman pointed out that this will give us no new knowledge. We must try to understand "…for the sake of what these transmissions are to be made…" – either by us or by ETIs…

## 2. Universality of consciousness?

How universal is consciousness? So far we have been lacking relevant experimental data. Only a single measurement – terrestrial realization of consciousness – is available. The aim of SETI is to try to find out whether consciousness is universal or not. A full description of the Universe as discussed by Linde (2003) –

*"Is it possible that consciousness, like space-time, has its own intrinsic degrees of freedom and that neglecting these will lead to a description of the universe that is fundamentally incomplete?"* –

is so far impossible to achieve – we do not know how to fit consciousness into the description of the Universe – as something unique, or as a universal phenomenon.

And it is not inconceivable that no one in the entire Universe knows this – the Universe is silent and even if there are other lone centers of consciousness somewhere else (Grinspoon 2003), THEIR physicists should face the same problem – how to fit consciousness into the description of the Universe – as a singular or a universal phenomenon. In this sense, the task of METI is to try to answer the question whether consciousness is universal – and this answer is to be meant for OTHERS…

Similarly, the Participatory Anthropic Principle (PAP) formulated by John Wheeler in 1983 – *"Observers are necessary to bring the Universe into being"* – is incomplete in the sense that the Universe that we now observe is a Silent Universe, a Universe of observers, whereas true participation in the scene of the Universe cannot be limited to mere contemplation.

One can speak about true "participation" when this "participation" becomes OBSERVABLE by a distant observer. Wheeler's Participatory Anthropic Principle should therefore be supplemented by the following statement:

*"Senders are necessary to bring consciousness into the Universe".*

So, the participation of senders would transform the observer's consciousness of the Universe into a consciousness that recognizes a Universe that is inhabited by at least two, separate intelligences (e.g., two civilizations). In turn, this transformation of the observer's consciousness would itself represent a contribution to existence. In other words, from an ontological perspective, senders would help observers better understand the true nature of being (assuming, of course, that the Universe is inhabited), and, in the process, change the very nature of being, i.e., into a state where the existence of extraterrestrial life is confirmed.

## 3. The Drake equation with the METI coefficient

The classic Drake equation is the product of seven parameters that estimate the number of potentially detectable extraterrestrial civilizations in our Galaxy:

$$N = R^* \times f_p \times n_e \times f_l \times f_i \times f_c \times L,$$

where $N$ = the number of potentially detectable civilizations in the Milky Way Galaxy; $R^*$ = the rate of formation of stars in the Galaxy; $f_p$ = the fraction of those stars with planetary systems; $n_e$ = the number of planets per solar system that are suitable for life; $f_l$ = the fraction of those planets where life actually appears; $f_i$ = the fraction of life sites where intelligence develops; $f_c$ = the fraction of communicative planets (those on which electromagnetic communications technology develops); $L$ = the "lifetime" over which such civilizations transmit detectable signals into space.

This equation takes into account many factors, but not all. Namely, it leaves out the fraction of emitting "intelligent planets," i.e., planets that are, like our Earth, in the communicative phase of their existence, and at the same time "bring" consciousness into the Universe by purposefully transmitting intelligent signals to the outside world. Estimation of this fraction is by no means just a question of idle curiosity given the attitude of our planetary consciousness toward such "bringing."

Here we are speaking about METI-phobia. It appeared immediately after the first interstellar radio message had been sent from Arecibo on November 16, 1974. Nobel Laureate Martin Ryle then published a protest where he warned: "…any creatures out there may be malevolent or hungry…" and called for an international ban to be imposed on any attempts to establish Contact and transmit messages from the Earth to hypothetical ETIs.

The International Academy of Astronautics (IAA) then adopted a Declaration (1989) calling for the restriction of such activities. Thus, paragraph 8 of this Declaration states: "No response to a signal or other evidence of extraterrestrial intelligence should be sent until appropriate international consultations have taken place. The procedures for such consultations will be the subject of a separate agreement, declaration or arrangement."

Six years later, the SETI Permanent Study Group of the IAA presented a Draft Declaration (1995), which envisages that a decision on whether or not to send an interstellar message should be approved by the United Nations General Assembly. Some researches operate with concepts of "peaceful civilization" and "aggressive civilization" and suggest that we should reply only to signals coming from a peaceful civilization – an attitude that would ultimately result in the total refusal to emit any signal at all. The reason: a message from a peaceful extraterrestrial civilization to which we are allowed to answer is impossible to distinguish from a message from an aggressive, but self-coding civilization, to which we should not reply. And given that we will be hardly able to develop an undoubted criterion to judge the altruism of the extraterrestrial civilization that would satisfy all those who fear the possible negative consequences of communicating, it would also be impossible to not only initiate, but even reply to interstellar messages. Our civilization would be doomed to eternal silence.

Unlike the English-language press, which has been discussing METI-phobia continuously, articles on this subject appear rarely in the Russian media. One of the most recent international campaigns involves a series of articles posted on the site of the SETI League and the adoption of the so-called "San Marino Scale" at the conference "We and SETI" held in San Marino in 2005. This scale, like the Richter scale for earthquakes, is meant to rank in-terstellar radio messages to ETI by the degree of risk. However, the Richter scale assesses real earthquakes that have already happened, whereas the San Marino scale assesses hypothetical, far-fetched consequences. In this context, of particular interest is the opinion of such fears and bans expressed by Paul Shuch, the SETI League's Executive Director: In 1998 he gave the following answer to our Internet poll, which we conducted during the period leading up to the Cosmic Call 1999 interstellar radio transmission: *"I am not an adherent of such isolationist (read paranoid) philosophy"*.

Our understanding of this problem stems from certain "double standards" (not in the common, negative meaning of this word combination): People fear that Something superpowerful and aggressive – such as the evil empires found in such modern, mythological/science fiction tales as the "Star Wars" serials – are either already aware of us, or will inevitably become aware of us. In this view, there is no escape from this fate. They will find us, first and foremost, by radio emission of dozens of military radars of USA and Russia, which are at the core of the national missile attack warning systems, which have been operating continuously 24 hours a day since the early 1970s (Morozov 2005). We must press forward Contact with all conceivable civilizations like our own, which being located far apart, may interact only by transmitting and receiving electromagnetic signals. And moreover, to be detected, we must emit targeted and guided messages toward the chosen celestial body.

However, we must take METI-phobia of extraterrestrial civilizations into account because of the current realities in Earth's civilization. To this end, the Drake equation should supplemented by the METI-coefficient $f_m$ (Zaitsev 2005):

$$N = R^* \times f_p \times n_e \times f_l \times f_i \times f_c \times f_m \times L,$$

where $f_m$ – the fraction of communicative civilizations (METI-civilizations), i.e., civilizations with clearly nonparanoidal planetary consciousness, which indeed produce planned and targeted interstellar messages. As mentioned above, to be in a communicative phase and emit METI messages is not the same thing. For example, we, although being in a communicative phase, are not a communicative civilization: We do not practice such activities as the purposeful and regular transmission of interstellar messages.

We may try to estimate the METI-coefficient $f_m$ for the only known, terrestrial civilization. As we pointed out above, our civilization is indeed in the communicative phase and it indeed conducts SETI activities. However, our METI/SETI ratio is less then one percent: these data follow from the review of Jill Tarter published in the recently released "SETI-2020" collection of papers (Tarter 2003). It lists 100 various SETI programs starting from the first OZMA project to our time. The total time of search is several years, whereas the total transmission time is only 37 hours (Zaitsev 2006). This characterizes the attitude of researches. However, we must also take into account the METI-phobia inherent to the planetary consciousness as a whole. And therefore if we assess the $f_m$ coefficient based on the only known civilization (and we are hardly peculiar if we are not alone), we find that it tends to zero and, consequently, the same should be true for the number of potentially detectable extraterrestrial civilizations. Hence, **the SETI Paradox**: *"Searching is meaningless if no one feels the need to transmit…"*

In other words: *"SETI makes sense only in a Universe with such properties that it develops Intelligence that realizes the need not only to conduct searches, but also to transmit intelligent signals to other hypothetical sites of self-consciousness"*.

It would become possible to establish Contact if one of the distinguishing features of Intelligence in our Universe is the missionary need to carry to Aliens the Good News that they are not alone in space. Given such enormous distances and, consequently, long signal propagation time, communications should be mostly one-way – our addressees receive our messages, and we, in turn, detect those who have chosen us as their addressees. This is how the Universe at a certain stage of its development appears for observers as inhabitable. Otherwise, centers of intelligence are doomed to remain lonely, unobserved civilizations.

And in conclusion, let us return to the beginning and give the classic quotation from the paper by Cocconi and Morrison (1959): *"The probability of success is difficult to estimate, but if we never search the chance of success is zero"*.

The above argument is, of course, true. However, accidental detection as a result of routine astronomical observations is also possible. However, this may happen only if there exist extraterrestrial civilizations that actually send interstellar messages. Therefore, in this context the above argument may be somewhat reformulated: *"The probability of success is difficult to estimate, but if nobody transmits the chance of success is zero in principle"*.

And we can formulate the following thesis implied by the SETI Paradox: *"Solely that who is overcoming the Great Silence deserves to hear the voice of the Universe"*.

-----

I am grateful to Richard Braastad for his valuable comments on the manuscript.